\documentclass[letterpaper]{article} 
\usepackage{aaai23}  
\usepackage{times}  
\usepackage{helvet}  
\usepackage{courier}  
\usepackage[hyphens]{url}  
\usepackage{graphicx} 
\usepackage{natbib}  
\usepackage{caption} 
\usepackage{algorithm}
\usepackage{algorithmic}

\usepackage{newfloat}
\usepackage{listings}

\usepackage{mathrsfs}
\usepackage{amsmath}
\usepackage{txfonts}
\usepackage{booktabs}
\usepackage{subfigure}
\usepackage{float}
\usepackage{pifont}
\usepackage[switch]{lineno} 

\DeclareCaptionStyle{ruled}{labelfont=normalfont,labelsep=colon,strut=off} 
\floatstyle{ruled}
\newfloat{listing}{tb}{lst}{}
\floatname{listing}{Listing}
\title{CLGT: A Graph Transformer for Student Performance Prediction \\in Collaborative Learning}
\author{
    Tianhao Peng\textsuperscript{\rm 1,\rm2},
    Yu Liang\textsuperscript{\rm 3}\thanks{Corresponding author},
    Wenjun Wu\textsuperscript{\rm 1,\rm 4},
    Jian Ren\textsuperscript{\rm 1,\rm2},
    Zhao Pengrui\textsuperscript{\rm 1,\rm2},
    Yanjun Pu\textsuperscript{\rm 1,\rm2}
}
\affiliations{
    \textsuperscript{\rm 1}State Key Laboratory of Software Development Environment, Beihang University\\
    \textsuperscript{\rm 2}School of Computer Science and Engineering, Beihang University\\
    \textsuperscript{\rm 3}Beijing Engineering Research Center for IoT Software and Systems, Beijing University of Technology\\
    \textsuperscript{\rm 4}Institute of Artificial Intelligence, Beihang University\\
    \{pengtianhao,wwj09315,renjian,zhaopengrui,buaapyj\}@buaa.edu.cn, yuliang@bjut.edu.cn
}

\usepackage{bibentry}

\begin{document}

\maketitle

\begin{abstract}
Modeling and predicting the performance of students in collaborative learning paradigms is an important task. Most of the research presented in literature regarding collaborative learning focuses on the discussion forums and social learning networks. There are only a few works that investigate how students interact with each other in team projects and how such interactions affect their academic performance. In order to bridge this gap, we choose a software engineering course as the study subject. The students who participate in a software engineering course are required to team up and complete a software project together. In this work, we construct an interaction graph based on the activities of students grouped in various teams. Based on this student interaction graph, we present an extended graph transformer framework for collaborative learning (CLGT) for evaluating and predicting the performance of students. Moreover, the proposed CLGT contains an interpretation module that explains the prediction results and visualizes the student interaction patterns. The experimental results confirm that the proposed CLGT outperforms the baseline models in terms of performing predictions based on the real-world datasets. Moreover, the proposed CLGT differentiates the students with poor performance in the collaborative learning paradigm and gives teachers early warnings, so that appropriate assistance can be provided.
\end{abstract}

\section{Introduction}\label{sec1}
In modern world, the collaborative learning (CL) is a prevalent learning method. In a CL environment, the students of different calibers and intellectual levels work together in teams and engage in a common task~\cite{laal2012collaborative}. The process of modeling and predicting the performance of students is an important task in CL. Software engineering (SE) is an educational program that combines theory and practice. The practical teaching forms the core of SE program. As the practical teaching includes considerable CL, we choose SE as the subject in this work.

The SE courses often incorporate team-based collaborative projects that are designed to mimic the professional software development tasks. These tasks are often more complex as compared to individual coding exercises implemented in other programming classes. In these tasks, each student in the study group needs to participate and work together on a medium scale software project following the classic life cycle of SE, following requirement specifications, architectural design, and coding. The final score of each student depends on their contribution in the project and the quality of the final software artifacts delivered by the team. In order to accurately and efficiently evaluate the academic performance of each student in a team project, the instructor needs an intelligent educational tool for analyzing the process of every project by mining the students’ behavior data, especially the interaction behaviors.

However, it remains a challenge for the instructors in the field of SE to score the team members by considering their individual contributions in the projects. This is mainly due to the difficulty in capturing and quantifying the amount of individual effort and work~\cite{parizi2018measuring}. A typical solution is to adopt either peer evaluation or assessing the team as a whole. However, both approaches can be inaccurate and prejudiced, resulting in an unfair assessment. Currently, with the introduction of online CL platforms, such as GitHub and GitLab, it is possible to collect the behavioral data of students, such as source code submissions and review postings. The availability of online behavior data enables the researchers to apply the machine-learning algorithms for CL~\cite{olsen2015predicting,yee2016predicting,ekuban2020using,yee2016stimulating}. However, these methods are mostly developed based on the classic machine learning models that choose the activity frequency of students as input features. Such coarse-granular features are unable to capture the rich spatial and temporal information embedded in the interaction of different team members, thus only presenting a limited value to the instructors.

The CL data contains interactive activities with the graph structured features, where the nodes represent the students and the edges between two students represent the interactions between them. There are often different kinds of interactions among the students, such as submitting documents and source codes, as well as posting reviews and issues. Each kind of edge has multiple attributes and has a potential impact on the final grade of students. Thus, the graph-structured interactive behaviors are formulated as a weighted heterogeneous graph, which contains not only multiple types of edges, but also the additional information features regarding each type of edge. In order to effectively exploit the rich edge feature information, in this work, we propose a method for processing the CL data and an extended \underline{G}raph \underline{T}ransformer framework for \underline{C}ollaborative \underline{L}earning (CLGT). We implement the proposed model and compare its performance with Ada Boost~\cite{DBLP:journals/jcss/FreundS97}, relational graph convolutional network (R-GCN)~\cite{schlichtkrull2018modeling}, graph transformer network (GTN)~\cite{yun2019graph} and Graph Trans~\cite{dwivedi2020generalization} based on the same dataset for predicting the performance of the students. The experimental results show that the proposed model outperforms four other models.

In addition to performance prediction, an instructor also needs intelligent tutoring tools to infer the latent impact of student interactions on the process of collaborative software development. For instance, the online questioning-answering and code submissions are recorded on the CL platform, however, the implicit activities, such as getting inspiration from reading other’s documents and codes is difficult to measure. Moreover, it is noteworthy that most of the deep learning models are inherently designed as black boxes, which are not able to provide good explanations regarding their reasoning mechanisms. Both factors, including the unobserved activities and the nature of the model makes it difficult to present the explainable prediction results to the instructors of SE courses. Therefore, in this work, we build an interpretation model inspired by PGM-Explainer~\cite{vu2020pgm} to explain the prediction results of the proposed CLGT model. Based on the results of the interpretation model, we are able to analyze whether the students influence their peers during the learning process and to what extent.

The major contributions of this work are summarized below.
\begin{enumerate}
\item In this work, we extend the graph transformer model to develop the CLGT model for accurately predicting and assessing the performance of students in a project-based CL environment.
\item We build an interpretation module for explaining that which part of the graph structure data is responsible for the prediction results of the proposed CLGT framework.
\item To facilitate further research, we publish our dataset of students CL activities generated from an online project-based SE course. The relevant code and dataset are publicly available in the code repository~\footnote{\url{https://github.com/Tianhao-Peng/CLGT/}}. 
\end{enumerate}

\begin{figure*}[htb]
\center{\includegraphics[width=0.75\textwidth]{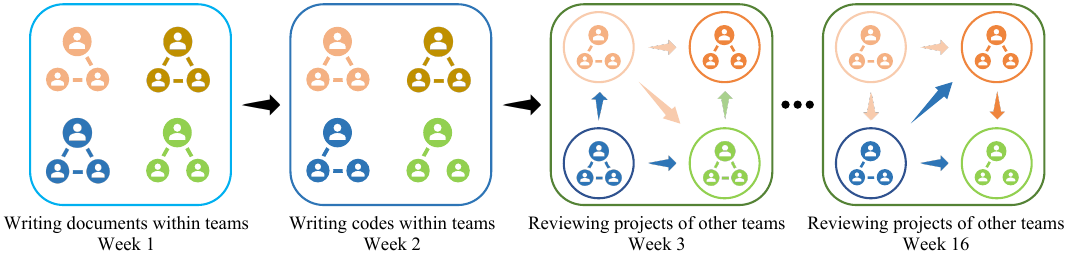}}
\caption{The flowchart of the three cycle phases of the SE course. During the documentation and coding phase, the students interact with other members of their teams, and during the reviewing phase, the integration occurs between the teams as well. As shown in the figure, the interaction graphs within and between the teams may change on weekly basis.}
\label{fig:course}
\end{figure*}

\section{Related Works}\label{sec2}

There are some works presented in literature for measuring the contribution of each individual in the projects related to software engineering courses. The log data and activity records available in version control systems are used in collaborative software development~\cite{arcelli2017students,loeliger2012version,clifton2007subverting,reid2005learning}, thus enabling the researchers to gather data regarding the computing contribution metrics. Most of the methods presented in literature~\cite{parizi2018measuring,buffardi2020assessing} utilize classic statistics and regression algorithms for assessing the contribution of a member based on the activity ratio among the metrics of team members, including the number of commits, pull requests, and the lines of code changed by this member. However, these methods fail to consider and reflect the interactive activities that occur during the software development process. Fu et al.~\cite{fu2021understanding} constructed a developer interaction graph based on the data obtained from GitHub (a git repository manager) for identifying the key developers bridging messages across the GitHub communities. This method only studies the structural properties of large-scale interaction networks in the professional software development communities without any attention to the topics of small- or medium-sized student interaction networks in collaborative learning. To the best of our knowledge, the proposed work is a first attempt that focuses on predicting the performance of students based on their interactive graph in software engineering courses.

It is noteworthy that most of the research efforts regarding the learning outcome prediction have focused on modeling the performance of students based on individual learning process~\cite{corbett1994knowledge,pavlik2009performance,piech2015deep,pu2019atc,liang2022help}. However, the individual analytic models cannot be applied directly in collaborative learning environments since they do not consider the influence of the students within a team or between different teams. There are few works that make an effort to extend the models designed specifically for individual learning to team-based learning scenarios. Olsen et al. ~\cite{olsen2015predicting} extended the additive factors model~\cite{pavlik2009performance} by using cooperative features to predict the performance of students in a collaborative learning environment. Yee-King et al. ~\cite{yee2016predicting} employed K-nearest neighbour (KNN) model to make predictions on the grades of students in a collaborative learning environment. Ekuban et al. ~\cite{ekuban2020using} evaluated multiple machine learning models, including decision tree, random forest, extra trees, Ada Boost, and gradient boosting to predict the performance of students working in a team based on student interactions. However, the models discussed above are shallow and have a limited performance in terms of handling large scale datasets and modeling complex graph-structured data based on spatio-temporal features.

The graph neural network (GNN) is a suitable framework for modeling complex interactions. The applications of GNN are widespread, ranging from image classification and video processing to speech recognition and natural language processing~\cite{wu2020comprehensive}. However, there are few research works that use GNN for modeling student interactions in a collaborative learning environment. The R-GCN~\cite{schlichtkrull2018modeling} proposes relation-specific transformation in the message passing steps to deal with various relations in edges. GTN~\cite{yun2019graph} proposes a novel graph transformer layer to identify the connections between unconnected nodes that are closely related. Graph Trans~\cite{dwivedi2020generalization} proposes an elegant positional encoding strategy based on the eigenvectors of the graph Laplacian to apply attention to neighbouring nodes. However, these models cannot fully and efficiently utilize the extra edge information of weighted heterogeneous graphs in some datasets which contain not only multiple types of edges, but also additional information features of each type of edge. 

Moreover, there has been little work devoted to explaining collaborative learning activities with explainable models. Without reasoning the underlying mechanisms behind the predictions, deep models cannot be fully used in collaborative learning. PGM-Explainer~\cite{vu2020pgm} is an explanation method that explains the predictions of any GNN in an interpretable manner. It has great performance in interpreting graph-structured data, including node prediction tasks. 

\begin{figure*}[htb]
\center{\includegraphics[width=0.75\textwidth]{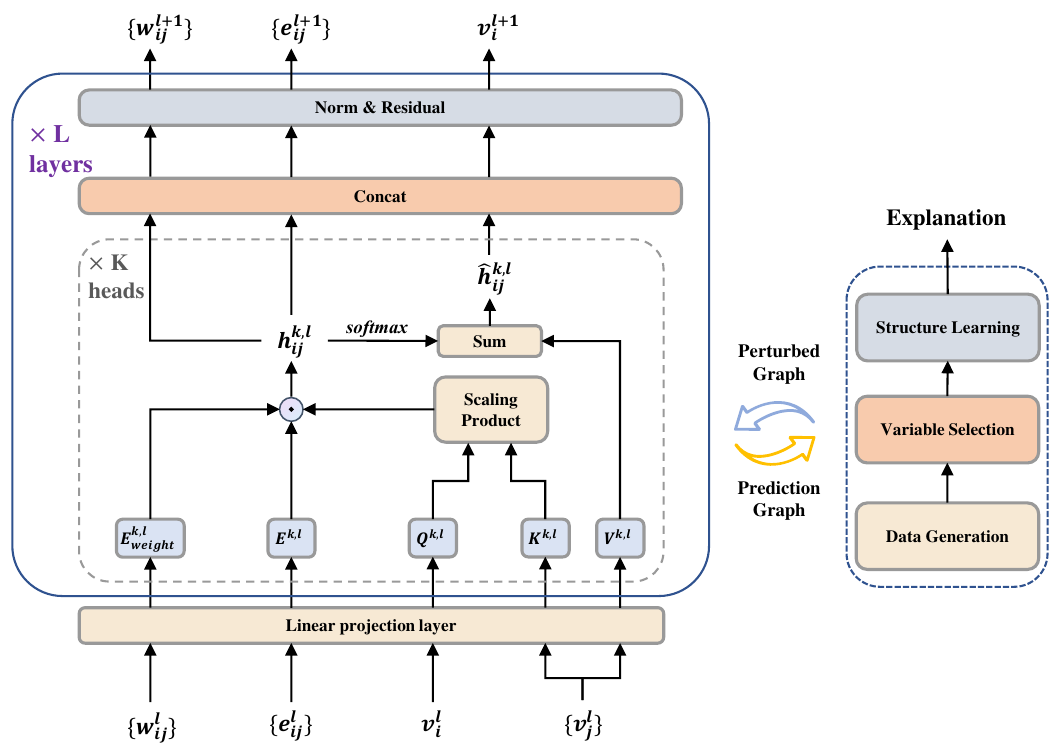}}
\caption{The architecture for the CLGT framework, including the prediction module (Left) and the explainer module (Right). }
\label{fig:transformer}
\end{figure*}

\section{Problem Statement}
In this work, we aim to model the performance of students during the implementation of SE projects for exploring their interactions with each other team members, and the influence of these interactions on the academic performance of students in a CL environment.

The goal of any software engineering course is to enable the students to master a variety of software engineering skills. The learning activities designed to achieve these learning goals are divided into a three-phase cycle, including writing documents, coding, and reviewing projects of other teams, as presented in Fig.~\ref{fig:course}. During the documentation phase, the students are required to submit various files. Each change in these files can be tracked by using a version control system. During the coding phase, the students commit code on the course website. Finally, during the reviewing phase, the students review the documents and codes of other teams and point out the problems by raising issues on the course website. Each problem raised by students has a degree rating, which indicates the severity of that problem. When students commit the revised documents or code in the repository, the rest of the team can view these changes and continue their development by considering these changes. Therefore, we assume that the commit behavior affects all the other members of the team. When a student raises a problem on a project of another team by creating an issue, all the members of that team have to work together to fix the project based on the raised issue. Therefore, we believe that the behavior of issue affects all the members of the team.

As presented in Fig.~\ref{fig:course}, the collaborative learning activities have peculiar spatial and temporal characteristics. In terms of spatial characteristics, the interaction activities reflect the structural influence relationships among the students in the course. In terms of temporal characteristics, the interaction graph keeps changing as the course progresses. The machine learning models, such as Ada Boost, KNN, and decision trees, currently used in collaborative learning research~\cite{olsen2015predicting,pavlik2009performance,yee2016predicting,ekuban2020using} are unable to deal with such complex spatio-temporal data. The GNNs are deep learning models that have the ability to capture the graph dependencies based on message passing between graph nodes. The transformers have the ability to forecast time series data~\cite{min2022transformer}. In this work, we combine GNN and Transformer models to make full use of the spatio-temporal features available in the data. We propose a new model CLGT based on Graph Trans ~\cite{dwivedi2020generalization} for modeling the interactions between the students participating in any software engineering course.

\subsection{Interaction Graph Generation}
It is necessary to convert the data collected from the version control system into a format that is usable by the neural networks. The students use a CL platform as a repository for code and document projects. The CL platform records the details regarding each change a student makes in any file. Therefore, the nature and place of the change are known. The data type of the interaction graph can be generated based on the recorded information. In this work, we require a definition that can validate if a construct is an instance of an interaction graph. We identify three necessary and sufficient conditions for interaction graphs.

\begin{enumerate}
\item An interaction graph is described by a graph $G = (V, E)$, where $V$ denotes a set of vertices and $E$ denotes a set of edges such that $E \in V \times V$. All the edges are directed from one vertex to another. This indicates that one vertex interacts with other vertices.
\item Each vertex $V$ represents a student, and each student is the part of a team.
\item Edges $E$ represents the interactions between different vertexes $V$. There are two types of edges, including interactions between students within the same team and interactions between different teams. Additionally, each edge type has different weights, depicting the degree of influence of the interaction activity.

\end{enumerate}

Since there are multiple types of edges in an interaction graph and each edge type has multiple types of attributes, an interaction graph is a weighted heterogeneous graph. 

\subsection{Interaction Degree Definitions}
In order to accurately depict the interaction activities between the students, we consider the revised lines in the documents and the codes. The issue depict the degree of influence of an interaction activity. The definition of the interaction degree of documents, codes, and issues is as follows.

\begin{enumerate}
\item For the interaction activities of document revision $a_{doc}$, the version control system record the number of lines added or deleted in the document $num_{doc}$, which indicates the degree of influence of the revision activity. Since the document revision activity is available to the students within the same team, we normalize the number of lines $num_{doc}$ between all commits performed in one week, as follows: 
$$
I(a_{doc}^i) = \frac{num_{doc}^i}{\sum_{k=0}^n num_{doc}^k}
$$
where $I(a_{doc}^i)$ indicates the influence of the document revision activity $a_{doc}^i$.

\item As the interaction activities regarding code revision $a_{code}$ are available to the students within the same team, we process the data similar to the previous step. We normalize the number of lines $num_{code}$ between all the commits performed in one week, as follows:
$$
I(a_{code}^i) = \frac{num_{code}^i}{\sum_{k=0}^n num_{code}^k}
$$
where $I(a_{code}^i)$ denotes the influence of the code revision activity $a_{code}^i$.

\item For the interaction activities regarding the creation of issues $a_{issue}$, each issue records the severity of the problems raised in the issue. We normalize the severity of the issues $num_{issue}$ between all issues incorporated in one week, as follows:
$$
I(a_{issue}^i) = \frac{num_{issue}^i}{\sum_{k=0}^n num_{issue}^k}
$$
where $I(a_{issue}^i)$ denotes the influence of the code revision activity $a_{issue}^i$. 
\end{enumerate}

Based on the normalized data of the document revision, code revision, and issue related activities, we generate the interaction matrix, which is used as the input of the model.

\subsection{Interaction Matrix Generation}
Once an interaction graph is generated based on the collected commit records and issue data, it can be converted into an interaction matrix. For a simple graph with vertex set $V = {v_1, \cdots, v_n}$, the interaction matrix $A$ is a square matrix of dimension $n \times n$. The value of $A_{ij}$ represents whether vertex $v_i$ interacts with $v_j$. Zero denotes that there is no interaction, while a positive number indicates the influence of the interaction. Since the interaction graph is directed, the interaction matrix is asymmetric. Especially, the diagonal elements of a matrix are zero, as the edges (cycles) from the vertices pointing towards themselves are not allowed in the interaction graphs.

\subsection{Prediction Module}
The proposed prediction module of CLGT framework is inspired by the Graph Trans~\cite{dwivedi2020generalization}.It is designed to effectively utilize the rich edge feature information in collaborative learning. It is notable that original graph trans model only considers heterogeneous graphs that contain multiple types of edges and does not consider the case where the weighted heterogeneous graph contains additional information features regarding the type of each edge. For instance, in our software engineering course dataset, the edges represent different interactions, and each interaction activity has a varying degree of influence. In order to address this problem, we add a pipeline for the weighted edge features based on the original model for making full use of edge information available in the graph. Fig.~\ref{fig:transformer} (Left) illustrates the prediction module of CLGT framework. The prediction module comprises an extra edge feature pipeline and an extra weighted edge feature pipeline to effectively utilize the available edge information.

Given the interaction graph $G=(V, E)$, we pass the input node $v_i$ for each node $i$, edge features $e_{ij}$ and weighted edge features $w_{ij}$ for each edge between node $i$ and node $j$ via a linear projection layer to embed these to $d-$dimensional hidden features $v^0_i$, $e^0_{ij}$ and $w^0_{ij}$. After the input layer, the pipeline propagates edge attributes from one layer to another, and the layer update equations are as follows.

\begin{equation}
h_{i j}^{k, \ell}=\left(\frac{Q^{k, \ell} v_{i}^{\ell} \cdot K^{k, \ell} v_{j}^{\ell}}{\sqrt{d_{k}}}\right) \cdot E^{k, \ell} e_{i j}^{\ell} \cdot E_{weight}^{k, \ell} w_{i j}^{\ell}
\end{equation}
\begin{equation}
\hat{h}_{i j}^{k, \ell} =  \sum_{j \in \mathcal{N}_{i}} softmax(h_{i j}^{k, \ell}) V^{k, \ell} v_j^{\ell}
\end{equation}

where $Q^{k, \ell}, K^{k, \ell}, V^{k, \ell} \in \mathbb{R}^{d_k \times d}$ are trainable parameter matrices, $k$ denotes the number of attention heads, $\mathcal{N}_{i}$ denotes the neighbors of node $i$. The outputs of $h_{i j}^{k, \ell}$ and $\hat{h}_{i j}^{k, \ell}$ in attention heads are then concatenated and passed via residual connection layers as follows.

\begin{equation}
v_{i}^{\ell+1}=F_r(F_c(\hat{h}_{i j}^{k, \ell})+v_{i}^{\ell})
\end{equation}
\begin{equation}
e_{i}^{\ell+1}=F_r(F_c(h_{i j}^{k, \ell})+e_{i}^{\ell})
\end{equation}
\begin{equation}
w_{i}^{\ell+1}=F_r(F_c(h_{i j}^{k, \ell})+w_{i}^{\ell})
\end{equation}

where $v_{i}^{\ell+1}, e_{i}^{\ell+1}, w_{i}^{\ell+1}$ denotes the output of $\ell$ layer and the input of $\ell+1$ layer, $F_c$ denotes the concat operation, $F_r$ denotes the normalization and residual connection operation.

The task of the model is node classification, $e_{i}^{\ell}, w_{i}^{\ell}$ represent the intermediate results, and $v_{i}^{\ell}$ in each layer is passed to a fully connected layer to compute the prediction scores.

\subsection{Explainer Module}
In order to understand the underlying mechanisms behind the predictions, and present the explainable prediction results, we build an explainer module for the prediction module. This explainer module is inspired by the PGM-Explainer~\cite{vu2020pgm} and elaborates the process based on which the proposed CLGT makes predictions. Moreover, it also shows the part of the graph structure data that is responsible for the prediction results.

As presented in Fig.~\ref{fig:transformer} (Right), the workflow of the explainer module can be roughly divided into three stages.
\begin{enumerate}

\item \textbf{ Data Generation}
The explainer module repeatedly perturbs the nodes in the original graph data and feeds the perturbed graph data to the proposed CLGT for obtaining the sampled data. This sample data contains node information and prediction results.
\item \textbf{Variable Selection}
Based on the sampled data, a pairwise dependency test is used to form an approximate Markov blanket for the target node to reduce the computational overhead and obtain all the potential statistics of the target node.
\item \textbf{Structure Learning} 
The explainer module uses a hill-climbing algorithm for maximizing the Bayesian information criterion (BIC) score in order to obtain an explanatory Bayesian network. The explainer module generates a weighted graph with the same number of nodes as the original input graph. The edge shows the influence of one node on the prediction result of the other node.
\end{enumerate}

\begin{figure*}[htb] 
	\centering
	\subfigure[Visualization of the output of the explainer module]{
		\label{fig:influence_overview}
		\includegraphics[width=0.50\linewidth]{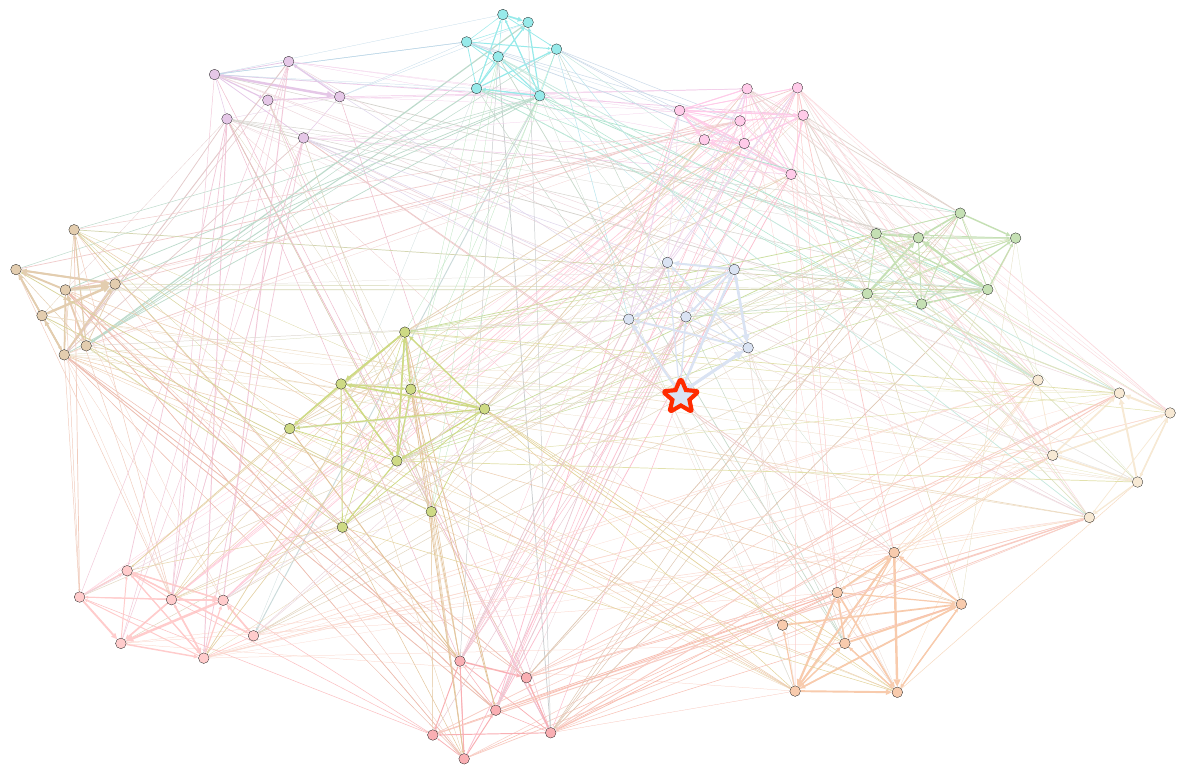}}
	\subfigure[Subgraph of Fig.~\ref{fig:influence_overview}]{
		\label{fig:one_student}
		\includegraphics[width=0.34\linewidth]{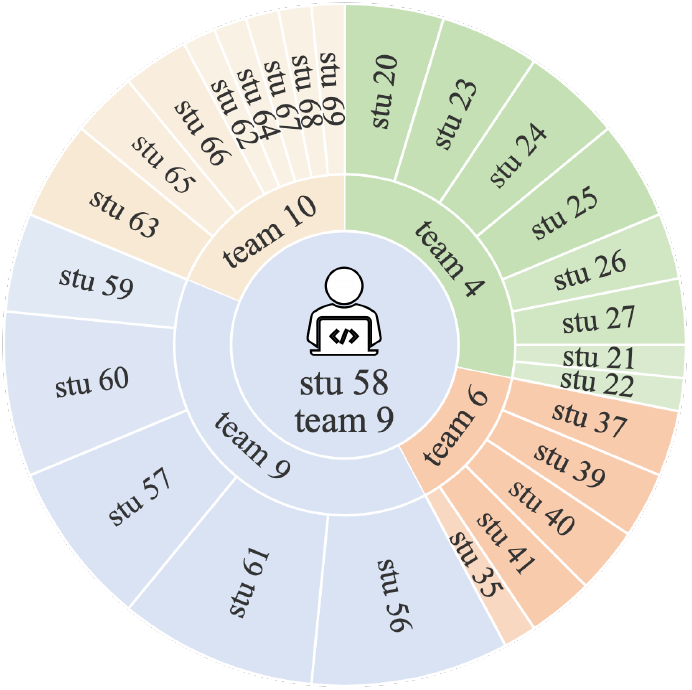}}
	\caption{(a) presents the visualization of the output of the explainer module. The nodes represent the students, and the edges represent the influence of students on each other. Different teams are denoted by different colors. The student 58 of team 9 is specially marked with \ding {80}, which is further analyzed in (b). (b) presents the visualization of part of the Fig.~\ref{fig:influence_overview}. Different colors in the inner circle represent different teams, and the outer circular sector represents different students of the corresponding teams. The proportion of the sector indicates the degree of influence of a team and its members on the middle student. The larger the proportion, the greater the influence.}
\end{figure*}

\section{Experiments}
\subsection{Dataset}
The object of this work include teams of students from a software engineering graduate course conducted in Spring 2021. In this course, a GitLab-based website~\footnote{\url{https://gitlab.com/}} is selected to implement the CL platform. We aim to study the interactions of students, when performing software engineering projects assigned in a semester to explore the ways students interact with each other in team projects and the effect of such interactions on the academic performance of students in a CL environment. The software engineering course lasted 16 weeks and comprised three different sessions, including writing documents, writing code, and reviewing projects of other teams. We acquire all the commits and issues available within the GitLab during the 16-week course. We obtain 4,903 commit records and 862 issues. Each commit records the number of lines added or deleted from the documents and code. Each issue records the severity of the problems raised in the issue. Each week, the teacher classified students' grades into three categories (A, B, C) based on the quality and quantity of documentation and code contributed in the software engineering project.

\subsection{Data Process}
Based on the proposed method, the interaction graphs and interaction matrices are easily generated. As the course lasted for 16 weeks and the teacher graded each student weekly, we divide the acquired data into 16 sections. In each section, we generate an interaction graph and three corresponding matrices, namely addition, deletion, and issue matrices. The addition matrix (or the deletion matrix) considers the members of the same team only and depict the influence of adding (or deleting) documents and codes on the members of the same team. The issue matrix considers the members of different teams and represents the influence of issues raised by one team on the members of other teams. Since the number of lines of code and documentation in each commit vary considerably, and there is a need for quantifying the severity of raised issues, we divide them into three levels, i.e., minor, moderate, and severe levels. After division, we obtain 48 matrices, i.e., three matrices per week. Now, each matrix consists of three types of elements, representing the influence from one vertex to another.

The purpose of the proposed CLGT is node classification. The labels of each node include the weekly and final grades of the students in the published dataset. The weekly and final grades of students are assigned by the teacher based on the quality and quantity of documentation and code contributed in the software engineering project.

\subsection{Baselines and Experimental Setup}
In the experiments, we compare the proposed CLGT with Ada Boost and three state-of-the-art graph neural networks. The CLGT, Ada Boost, R-GCN, GTN and Graph Trans are constructed using Pytorch~\cite{paszke2017automatic} and scikit-learn~\cite{conf:scikit-learn}. For Ada Boost model, we use \textit{sklearn.ensemble.AdaBoostClassifier} with 100 week estimators given the cumulative average number of GitLab pushes per week, and the total number of lines of documentation and code added, deleted, and modified per week. In the case of R-GCN and GTN models, the Adam optimizer is used and the hyperparameters, including learning rate, weight decay etc. are selected appropriately so that each baseline yields its best performance. For the original graph trans model and the proposed CLGT model, we use 10 graph transformer layers, where each layer comprises 8 attention heads and arbitrary hidden dimensions. Therefore, the total number of trainable parameters is in the range of 588k and 855k. We use the learning rate decay strategy for training the models. The training process is stopped when the learning rate reaches a value of $1 \times 10^{-6}$.

\section{Results and Discussion}
\subsection{Prediction Results}

\begin{figure*}[htb]
\center{\includegraphics[width=\linewidth]{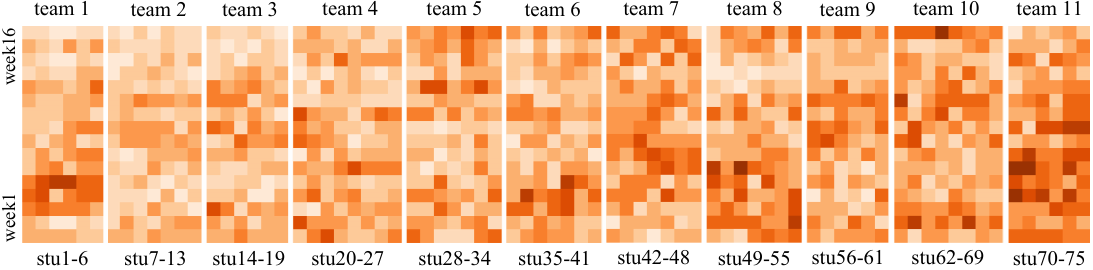}}
\caption{The horizontal axis represents all the 75 students in the course. The vertical axis represents the 16 weeks during which the course is conducted. The shades of color represent the activeness of students in the course, with darker colors representing more active students. Additionally, 75 students are divided into 11 teams, with each block in the figure representing a team.} 
\label{fig:all_students}
\end{figure*}

The experimental results are presented in Table~\ref{table:tb1}. For the sake of comparison, we implement the proposed CLGT model, R-GCN, GTN, Graph Trans model and Ada Boost on the same dataset we collected. In this work, the accuracy (ACC), the F1-score and the average and standard deviation of the area under the ROC curve (AUC) are used as evaluation metrics. The larger the AUC, F1-score or ACC, the better the model's prediction performance. 

Experimental results show that CLGT model performs better as compared to Ada Boost, R-GCN, GTN, and Graph Trans model in terms of ACC, F1-score, and AUC. The outstanding performance of the proposed model shows better utilization of rich edge and weighted edge feature information as compared to the baseline models. The proposed architecture specifically ensures better utilization of datasets that not only contain multiple types of edges, but also additional information features regarding the type of each edge.

\begin{table}[htb]
    \renewcommand\arraystretch{1.2}
    \centering
    \small
    \scalebox{1.0}{
    \begin{tabular}{lccc}
    \toprule
    \textbf{Method} & \textbf{ACC} & \textbf{F1-score} & \textbf{AUC} \\
    \midrule
    Ada Boost & 41.53$\pm$1.30  & 29.36$\pm$1.82  & 48.81$\pm$1.05  \\ 
    R-GCN & 64.81$\pm$2.13 & 52.64$\pm$1.83 & 76.54$\pm$2.60 \\
    GTN  & 73.37$\pm$2.32 & 62.79$\pm$2.25 & 78.84$\pm$1.12 \\
    Graph Trans &  72.46$\pm$2.08 &  59.04$\pm$1.84 &  86.21$\pm$1.73\\
    \midrule
    {\bfseries CLGT}(ours) & {\bfseries 73.92$\pm$1.78} & {\bfseries 66.75$\pm$2.15} & {\bfseries 90.57$\pm$1.88} \\
    \bottomrule
    \end{tabular}
    }
    \caption{The prediction performances of the proposed CLGT, Ada Boost, R-GCN, GTN, and Graph Trans model based on the same dataset. The best results are highlighted.}
    \label{table:tb1}
\end{table}

\subsection{Case Study and Visualization}
In order to explore the explainability of the predicted results obtained using the proposed CLGT model, we build an explainer module that generates a weighted graph. The edges of this weighted graph represent the influence of students on each other. The output of the weighted graph is visualized (as presented in Fig.~\ref{fig:influence_overview}) for performing intuitive analysis.

\begin{enumerate}
\item After considering a specific student, we explore which students have the greatest influence on him or her. In most cases, the teammates have a greater influence on the students as compared to other team members. However, we can also infer some information from the interpretation results that is not reflected directly in the original interaction graph. We process and visualize the subgraph of Fig.~\ref{fig:influence_overview} in Fig.~\ref{fig:one_student}. The student 58 (in team 9) does not interact with student 22 (in team 4). However, that student indirectly affects student 58 through the interaction with students in team 10 (there are direct interactions between team 4 and team 10). This shows that we can explore potential relationships among different students to further analyze the role of student behavior.

\item We explore how the influence of each student changes during the course of 16 weeks. We process the data of 16-week course in order to visualize the influence of all students in each week presented in Fig.~\ref{fig:all_students}. We find a high correlation between the students’ influence and the teams they belong to. Although some teams, e.g., team 10, do not achieve the best grades, their overall influence is significantly bigger. On the contrary, other teams, e.g., team 3, achieve better grades; however, their influence is not as high as their grades. We think this may reflect the underlying characteristics of different teams, such as team 10 being more active, but the overall quality of the project code is poor. On the other hand, team 3 showcases the exactly opposite characteristics. These results enable the course instructors to build better student profiles and perform accurate assessment based on the performance of students working in teams.
\end{enumerate}

\section{Conclusions}
In this work, we investigate how students interact with each other in team projects and how such interactions affect the academic performance of these students in the collaborative learning paradigm. We choose one-semester software engineering course to investigate and assess the collaborative learning activities of students. We propose a method for extracting the activity data from an in-house Gitlab platform and constructing the interaction graph. Moreover, we also introduce an extended graph transformer model named CLGT for accurately predicting the performance of students. We also build an interpretation model for explaining the prediction results of the proposed CLGT model and analyze the influence of each student on the teammates during the course project. The experimental results confirm that the proposed CLGT model outperforms the state-of-art models and provides good insights for the instructors to assess the learning progress of students. Moreover, the proposed CLGT differentiates the students with poor performance in collaborative learning and gives the teachers early warnings, so that appropriate assistance can be provided.

\section{Acknowledgments}
This work is supported in part by the Science and Technology Innovation 2030—“New Generation Artificial Intelligence” Major Project (2018AAA0102300) and the State Key Laboratory of Software Development Environment (SKLSDE-2020ZX-01/2022KF-08/2022KF-10).

\bibliography{aaai23}

\end{document}